\documentclass[12pt]{article}
\usepackage[a4paper,margin=1in,footskip=0.5cm]{geometry}
\usepackage{dcolumn}
\usepackage{changepage}
\usepackage{graphics}
\usepackage[pdftex]{graphicx,xcolor}
\usepackage{caption}
\usepackage{subcaption}
\usepackage{booktabs}
\usepackage{adjustbox}  
\usepackage{tabularx}
\newcolumntype{Y}{>{\centering\arraybackslash}X}
\usepackage{longtable} 
\usepackage{epstopdf}
\usepackage{mathtools,amsmath}
\usepackage{appendix}
\usepackage{lscape}
\usepackage{float}
\usepackage{hyperref}
\usepackage{authblk}
\usepackage{tabulary}
\usepackage{morefloats}
\usepackage{adjustbox,lipsum}
\usepackage[flushleft]{threeparttable}
\usepackage[utf8]{inputenc}
\usepackage[english]{babel}
\usepackage[autostyle]{csquotes} 
\usepackage{rotating}
\usepackage{placeins}
\usepackage{natbib,hyperref}
\usepackage{doi}
\usepackage{float}
\usepackage{flafter}
\usepackage{setspace}
\usepackage{textgreek}
\usepackage{grffile} 
\usepackage{array}
\usepackage{amsthm}
\usepackage{multirow}
\usepackage{tabularx}
\usepackage{lmodern}
\usepackage{pdflscape}
\DeclareUnicodeCharacter{00A0}{~} 
\usepackage{setspace}
\usepackage{comment}
\usepackage{dingbat}
\usepackage{chngcntr}

\setcitestyle{round}

\usepackage{natbib}

\title{The properties of a Leontief production technology for Health System Modeling: the \emph{Thanzi la Onse} model for Malawi}
\author{Martin Chalkley$^1$, Sakshi Mohan$^1$, Margherita Molaro$^2$, Bingling She$^2$ and Wiktoria Tafesse$^1$  \thanks{
1 Centre for Health Economics, University of York, UK
2 Department of Infectious Disease Epidemiology, School of Public Health, Imperial College London, London, UK
Corresponding author: Chalkley: Centre for Health Economics, University of York,\\ Martin.Chalkley@york.ac.uk.}}

\begin{document}
\onehalfspacing

\maketitle

\begin{abstract}

As health system modeling (HSM) advances to include more complete descriptions of the production of healthcare, it is important to establish a robust conceptual characterisation of the production process. For the \emph{Thanzi La Onse} model in Malawi we have incorporated an approach to production that is based on a form of Leontief technology -- fixed input proportions. At first sight, this form of technology appears restrictive relative to the general conception of a production \emph{function} employed in economics. In particular, the Leontief technology is associated with constant returns to scale, and level sets that are piecewise linear, both of which are highly restrictive properties. In this article we demonstrate that once incorporated into an all disease, agent-based model these properties are no longer present and the Leontief framework becomes a rich structure for describing healthcare production, and hence for examining the returns to health systems investments.

\end{abstract}

\newpage
\section{Introduction}\label{intro}
Health system models (HSMs) provide the means to analyse and evaluate health system interventions \citep{chang_dynamic_2017}. The Thanzi la Onse (TLO) model for Malawi \citep{hallett_estimates_2025} is a unique multi-disease whole system approach that provides a framework for simulating population health into the future accounting for the evolution of diseases and the impact of healthcare delivery. 

In developing this model we have been concerned with increasing the functionality of the elements that account for healthcare delivery and the link between available resources -- healthcare workers, consumables and equipment -- and the number of treatments that healthcare facilities can provide. This has led to developing a healthcare production framework in which we adopted fundamental principles from the economics of production \citep{rasmussen_production_2012}.

The concept of a \emph{production function} is crucial to implementing this approach within the TLO model, and there are many choices of such functions. However, the data requirements for paremeterising most of these functions cannot by met from existing sources. We therefore focused on a comparatively simple functional form -- the Leontief production function. 

In respect of the economic theory of production the Leontief function imposes very strong restrictions on the production process being modeled, including an assumption of constant returns to scale and the absense of substitution possibilies between inputs. The latter gives rise to (piecewise) linear level sets  which are termed isoquants in production theory. Put simply the Leontief function is a (mostly) linear representation of the relationship between inputs and outputs. That would seem very unlikely to capture what we observe to be non-linear responses between resources and treatments in the model.

However, as we explored the implementation of the Leontief function in the context of the highly multi-dimensional, multi-agent, dynamic and stochastic TLO model we discovered it is capable of capturing highly non-linear relationships between healthcare resources and treatments. This note sets out an explanation for how observable complexity can arise from a simple origin.

\section{The simple one output two input Leontief production function}\label{model}
We begin with the idea that the productive capacity of a healthcare system is determined by the resources it has at its disposal. In the simplest conception, we can consider a single measure of output -- the number of treatments that are possible and denoted by $y$ -- as depending on two measures of input which for convenience we refer to as amount of healthcare worker time, denoted $w$ and the volume of consumables, denoted $c$. 

The relationship between inputs and output is determined by what is termed the production \emph{technology} which can be conveniently summarised by a mathematical function: 
\begin{equation}
y=f(w,c)
\end{equation}
It is assumed that $f$, called the  \emph{production function}, is increasing in its arguments and quasi-concave, implying that its level sets, termed \emph{isoquants}, are convex curves (viewed from the origin). There is particular interest in what happens to $y$ as both inputs increase in equal proportions, this being the extent of \emph{returns to scale} (rts). If output increases in proportion with all inputs there are constant rts and decreasing (increasing) rts if output increases in less (greater) than proportion to inputs.

Figure 1 illustrates these properties graphically. In the left sub-figure the horizontal axis assumes all inputs increasing in proportion, and A shows how treatments increase if there constant rts, B shows the case of decreasing rts and C the case of increasing rts. In the right sub-figure the axes represent changing availability of the two inputs and the curves show combinations of those inputs that give rise to the same number of treatments.

\begin{figure}[htb]
\centering
\begin{subfigure}{.5\textwidth}
  \centering
  \includegraphics[width=.8\linewidth]{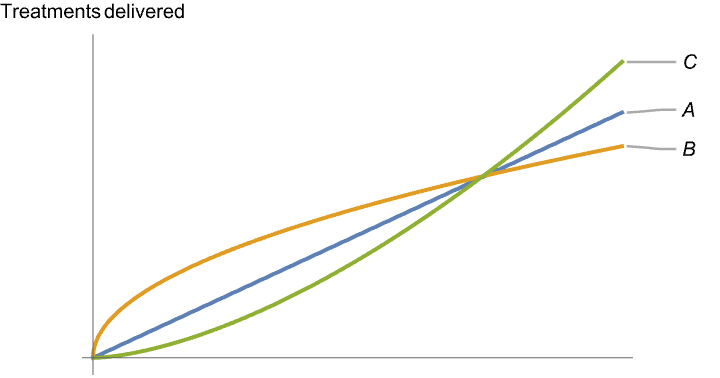}
  \caption{Returns to scale: Constant (A), Decreasing (B) and Increasing (C)}
  \label{fig:sub1}
\end{subfigure}%
\begin{subfigure}{.5\textwidth}
  \centering
  \includegraphics[width=.6\linewidth]{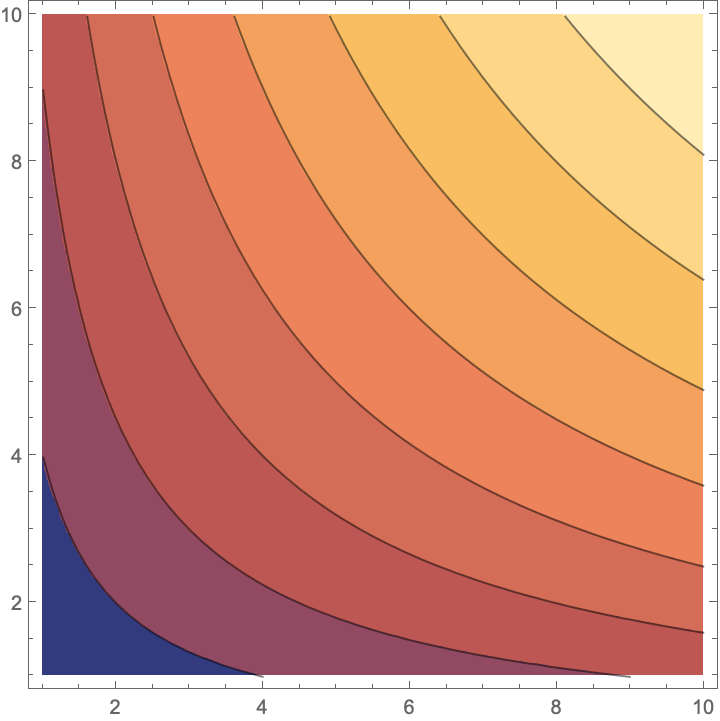}
  \caption{Isoquants with substitutability}
  \label{fig:sub2}
\end{subfigure}
\caption{Graphical properties of a typical production function}
\label{fig:test}
\end{figure}
In order to offer as flexible an approach as possible, functional forms for $f$ are specified to allow for variation in the shapes and positions of isoquants, and to allow for different returns to scale, both of which are determined through a discrete set of parameters. 

A number of popular functional forms have emerged as the basis for empirical examination of production relationships and one of the more general and frequently used of these is the \emph{constant elasticity of substitution} function \citep{solow_contribution_1956,arrow_capital-labor_1961,mcfadden_constant_1963} with a functional form of;
\begin{equation}
f(w,c) =F(aw^\rho+(1-a)^\rho)^\frac{v}{\rho}.
\end{equation}
This functional form has four parameters; $F$ (total factor productivity), $a$ (factor share parameter), $\rho$ (substitution parameter), and $v$ (return to scale parameter).

Although the CES function can accommodate subtle trade-offs between $w$ and $c$, the requirements to provide reasonable estimates of the parameters are very demanding. This is true even when there is a single output and two inputs, but it becomes extreme when as in the context of an all-diseases HSM, there are many possible outputs and many inputs to consider. For the TLO model for Malawi, there is no way to estimate these parameters given current or any conceivably likely future data.

A simpler conception of the relationship between input and output is to assume that each unit of output, i.e. each treatment, requires fixed amounts of healthcare worker time and consumables. This is the essence of a Leontief technology that we have adopted for the TLO model. In the two input, single output case this can be summarised in a a production function,
\begin{equation}
\label{leontief}
    y=\min\left(\frac{w}{a_1},\frac{c}{a_2}\right),
\end{equation}
where $a_1$ is the \emph{input requirement} of healthcare workers for each treatment. To calibrate the input requirements $a_1$ and $a_2$, there are a number of data sources that we can use, and these sources also relate to the different treatments represented in the model.

The feasibility of the Leontief formulation for an HSM model comes at the expense of flexibility in specifying production. The differences are summarized in Table \ref{tab:comparison} below and depicted graphically in the two panes of Figure 2.

\begin{table}[h]
    \centering
    \begin{tabular}{cc}
      \textbf{CES}   &  \textbf{Leontief}\\
 \hline
    Constant, decreasing or increasing returns to scale & Constant returns to scale\\
    Non-linear isoquants & Piecewise linear isoquants \\
    Substitution between inputs possible& No substitution possible
    \end{tabular}
    \caption{Characteristics of the CES and Leontief production functions}
    \label{tab:comparison}
\end{table}

\begin{figure}[htb]
\centering
\begin{subfigure}{.5\textwidth}
  \centering
  \includegraphics[width=.8\linewidth]{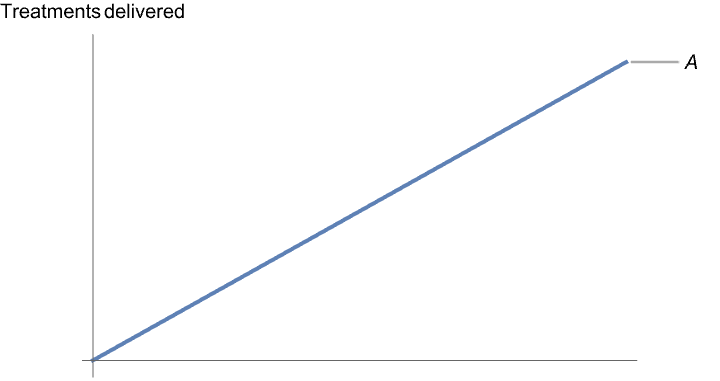}
  \caption{Constant returns to scale}
  \label{fig:sub3}
\end{subfigure}%
\begin{subfigure}{.5\textwidth}
  \centering
  \includegraphics[width=.6\linewidth]{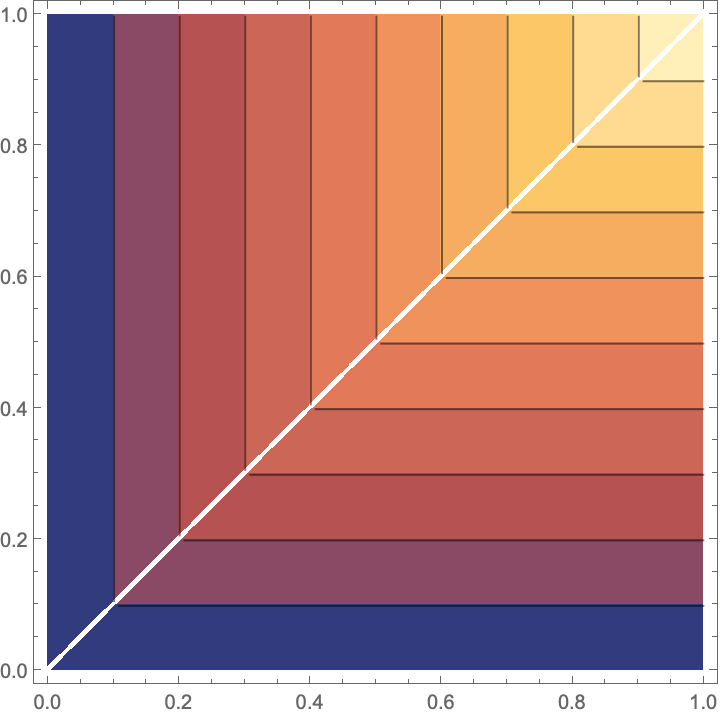}
  \caption{Piecewise linear isoquants}
  \label{fig:sub2}
\end{subfigure}
\caption{Graphical properties of a Leontief production function}
\label{fig:test}
\end{figure}
\section{Extension to two outputs and uncertainty} 
Assuming a Leontief production function would appear to greatly restrict the ability of an HSM to capture complex relationships between available resources and the treatment capacity of a health system. However, this fails to account for the complexity and richness of an agent-based dynamic model that incorporates multiple diseases. 

One key departure of the TLO model from the production functions envisioned above is the dimensionality of \emph{inputs}. Rather than one input for healthworker time $w$, the model allows for 9 different cadres of health workers. Similarly, the TLO model allows for many different drugs / consumables. However, increasing the number of inputs does not impact on the properties of the Leontief technology summarised in Table \ref{tab:comparison}.

A second key departure is that in the TLO model there are many different outputs - one for each of the different health system interactions considered in the model. These each require potentially different combinations of the many inputs considered. The multiplicity of outputs turns out to be crucial for the complexity of production relationships that the Leontief technology can capture.  To illustrate this we consider here just the simplest generalisation of adding a second output.

Viewed from the perspective of the first output  ($y_1$) the production of a second output ($y_2$) reduces the resources available. Assuming that $y_2$ has input requirements of $\frac{1}{b_1}$ and $\frac{1}{b_2}$ we can extend the definition of the production function for $y_1$ to account for this, as follows;

\begin{equation} \label{extendedleontief}
    y_1=\min\left(\frac{1}{a_1}\left(w-\frac{1}{b_1} y_2\right),\frac{1}{a_2}\left(c-\frac{1}{b_2}y_2\right)\right),
\end{equation}

For any given value of $y_2$ the function (\ref{extendedleontief}) is the same as that in (\ref{leontief}) with the values of $w$ and $c$ reduced by $\frac{1}{b_1}y_1$ and $\frac{1}{b_2}y_1$ respectively. It displays the same properties as (\ref{leontief}) except that as inputs $w$ and $c$ double $y_1$ will more than double and there are apparently increasing rts.  This is because the presence of $y_2$ acts as a kind of negative input. Constant rts is restored only if we consider a proportionate increase in all inputs \emph{and} $y_2$. Even at this most straightforward of generalisations indicates that the properties of the Leontief technology can become richer as we move towards its implementation in a multi-output model. Nevertheless, the shape of the isoquants generated by (\ref{extendedleontief}) are the same as those generated by (\ref{leontief}) albeit they are shifted by the presence of $y_2$.  Figure illustrates this.
\begin{figure}[htb]
\centering
\begin{subfigure}{.5\textwidth}
  \centering
  \includegraphics[width=.6\linewidth]{fig2a.png}
  \caption{With zero second output $y_2$}
  \label{fig:sub1}
\end{subfigure}%
\begin{subfigure}{.5\textwidth}
  \centering
  \includegraphics[width=.6\linewidth]{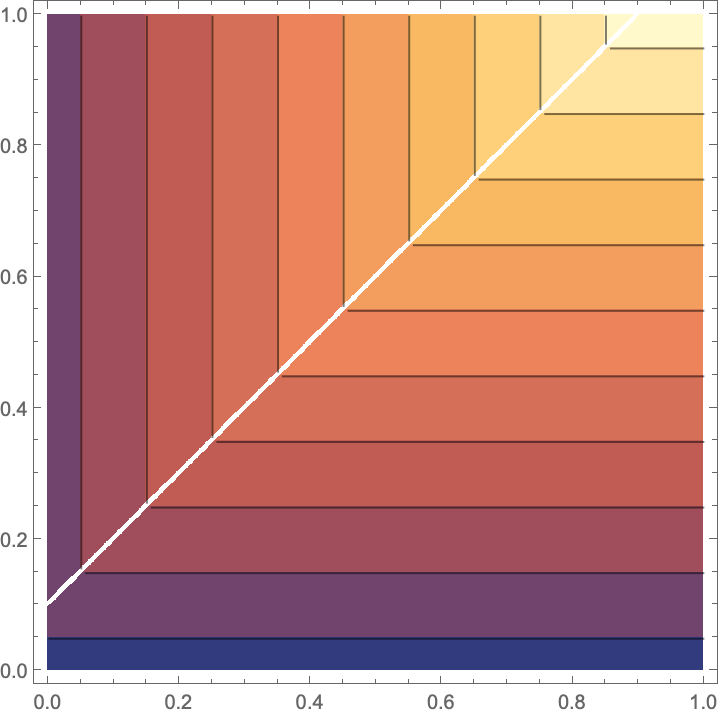}
  \caption{With positive second output $y_2$}
  \label{fig:sub2}
\end{subfigure}
\caption{Isoquants of a Leontief function with a second output $y_2$}
\label{fig:test}
\end{figure}

A third key departure in the TLO model is that the treatments required of the health system are generated through simulation of diseases at the level of individuals.  Thus, the resources consumed by $y_2$ in our example are determined probabilistically.  Viewed from the perspective of the ability of the health system to produce treatments $y_1$, the available resources are subject to uncertainty. This is a fundamental and important influence on the way that a Leontief technology can represent production possibilities.  

We can illustrate this by reference to the two output case. Consider (\ref{extendedleontief}) in which $y_2$ is a random variable with density $g(y_2)$. Over many iterations of the model, the observed average relationship between inputs $w$ and $c$, and the number of treatments $y_1$ produced will be given by
\begin{equation} \label{expectedleontief}
    E[y_1]=\int\min\left(\frac{1}{a_1}\left(w-\frac{1}{b_1} y_2\right),\frac{1}{a_2}\left(c-\frac{1}{b_2}y_2\right)\right)g(y_2)dy_2,
\end{equation}
where the integration is over the support of $y_2$.

\begin{figure}[htb]
\centering
\begin{subfigure}{.5\textwidth}
  \centering
  \includegraphics[width=.6\linewidth]{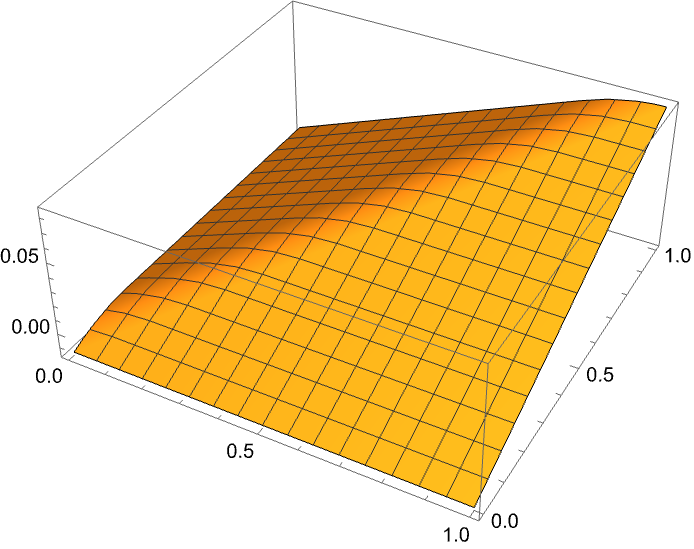}
  \caption{Leontief function with random second output}
  \label{fig:sub4a}
\end{subfigure}%
\begin{subfigure}{.5\textwidth}
  \centering
  \includegraphics[width=.6\linewidth]{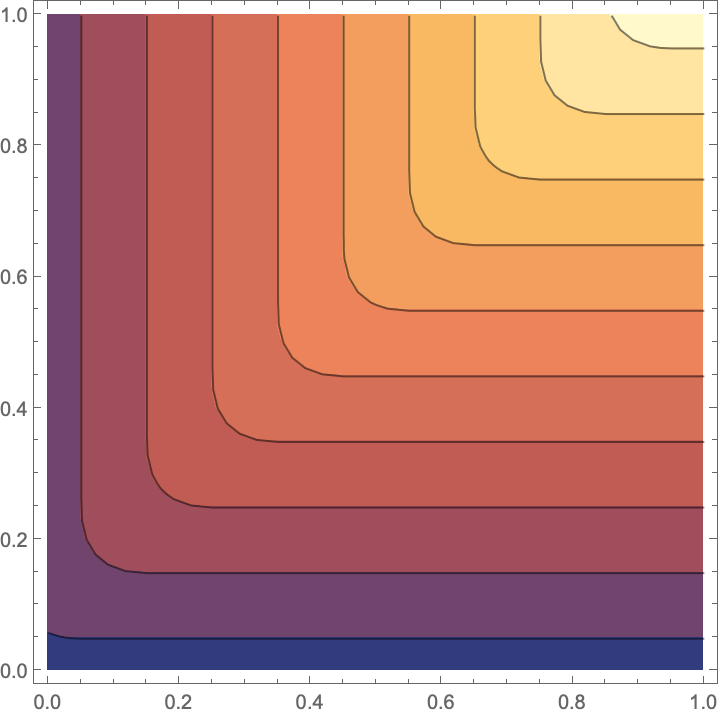}
  \caption{Isoquants of production function in (a)}
  \label{fig:sub4b}
\end{subfigure}
\caption{The multi-output Leontief production function for output $y_1$ where $y_2$ is determined probabilistically}
\label{figure4}
\end{figure}

The expected output in (\ref{expectedleontief}) is a function of $w$ and $c$ alone and is shown in \ref{fig:sub4a} and \ref{fig:sub4b}.
Hence, in its practical implementation in the multi-disease (and thus multi-treatment) TLO model, the Leontief production function gives rise to non-linear level sets and non-constant returns to scale analogous to the the properties of a more general production functions such as the CES.

\begin{figure}[htb]
\centering
\begin{subfigure}{.5\textwidth}
  \centering
  \includegraphics[width=.6\linewidth]{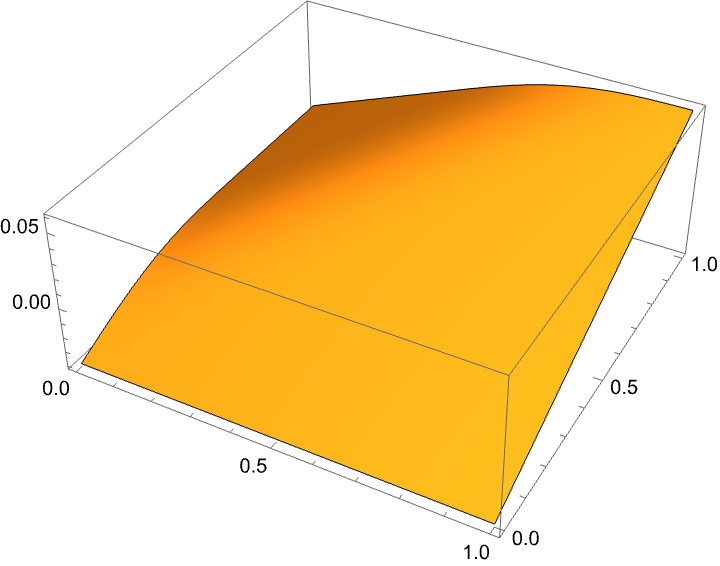}
  \caption{Leontief with 3 outputs}
  \label{sub5a}
\end{subfigure}%
\begin{subfigure}{.5\textwidth}
  \centering
  \includegraphics[width=.6\linewidth]{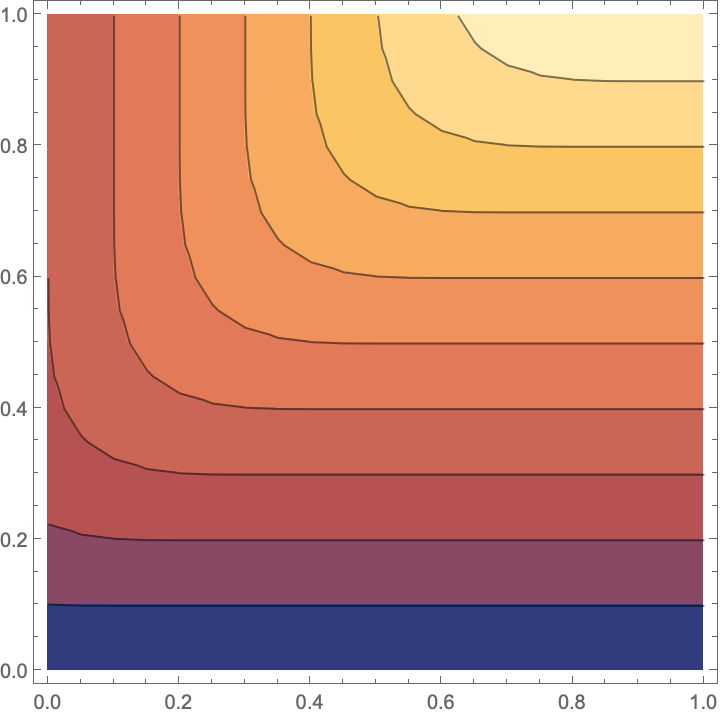}
  \caption{With positive second output $y_2$}
  \label{sub5b}
\end{subfigure}
\caption{Isoquants of production function in (a)}
\label{figure5}
\end{figure}

The specific form of the isoquants in \ref{figure4} is limited by having considered only a single additional output. In a truly multi-dimension output setting more general forms of naturally arise. For example, considering three outputs $y_1,y_2$ and $y_3$ and allowing for different degrees of correlation between the outputs $y_2$ and $y_3$, isoquants with a greater deviation from piecewise linearity can be generated.  This is illustrated in Figure \ref{figure5} .

\section{Conclusion}
In this article we have demonstrated that a Leontief production function that is implemented  into an all disease, agent-based health system  model becomes a rich structure for describing healthcare production, and hence for examining the returns to health systems investments. 

This is a potentially important insight because data to inform the parameters of more complex conceptualisation of production functions do not typically exist, and are unlikely to be forthcoming in the foreseeable future.  Our conclusion is that HSMs can nevertheless flourish and capture complex realities, by adopting a simple Leontief production function approach. 
\bibliographystyle{plainnat}
\bibliography{Leontiefbib}
\newpage
\section*{Technical notes}
Figures were created using Wolfram Mathematica and the supporting notebook containing the relevant code is accessible from \href{https://www.wolframcloud.com/obj/mjchalkley/Published/The%20properties%20of%20Leontief%20production%20function%20in%20the%20context%20of%20the%20TLO%20Model.nb}{Mathematica Notebook for Producing Figures}.

For simulating the Leontief function for one output subject to random variation in other outputs the Uniform [0,1] distribution was used throughout.

To account for potential correlation between other outputs a joint uniform distribution was used, either assuming independence or using the Ali-Mikhail-Haq copula to capture a general form of interdependence \citep{ali_class_1978}.

\end{document}